# Hierarchical Event Descriptor library schema for EEG data annotation

Dora Hermes[1,†,*], Tal Pal Attia[1,*], Sándor Beniczky[3,4], Jorge Bosch-Bayard[5], Arnaud Delorme[6], Brian Nils Lundstrom[7], Christine Rogers[8], Stefan Rampp[9,10], Seyed Yahya Shirazi[6], Dung Truong[6], Pedro Valdes-Sosa[5], Greg Worrell[7], Scott Makeig[6], Kay Robbins[2]

[1] Multimodal Neuroimaging Laboratory, Department of Physiology and Biomedical Engineering, Mayo Clinic, Rochester, Minnesota, USA
[2] Department of Computer Science, University of Texas San Antonio, San Antonio, Texas
[3] The Danish Epilepsy Centre, Filadelfia, Denmark
[4] Aarhus University and Aarhus University Hospital, Aarhus, Denmark
[5] Department of Biological and Health Psychology, Faculty of Psychology, Universidad Autónoma de Madrid, Spain
[6] Swartz Center for Computational Neuroscience, Institute for Neural Computation, University of California San Diego, La Jolla, California
[7] Bioelectronics, Neurophysiology, and Engineering Laboratory, Department of Neurology, Mayo Clinic, Rochester, Minnesota, USA
[8] McGill Centre for Integrative Neuroscience, Montreal Neurological Institute, Montreal, Canada
[9] Department of Neurosurgery, University Hospital Erlangen, Erlangen, Germany
[10] Department of Neurosurgery, University Hospital Halle (Saale), Halle (Saale), Germany

* Contributed equally
† Corresponding: Dora Hermes (hermes.dora@mayo.edu)

**Abstract**

Standardizing terminology to annotate electrophysiological events can improve both computational research and clinical care. Sharing data enriched with standard terms can facilitate data exploration, from case studies to mega-analyses. The machine readability of such electrophysiological event annotations is essential for performing analyses efficiently across software tools and packages. Hierarchical Event Descriptors (HED) provide a framework for describing events in neuroscience experiments. HED library schemas extend the standard HED schema vocabulary to include specialized vocabularies, such as standardized clinical terms for electrophysiological events. The Standardized Computer-based Organized Reporting of EEG (SCORE) defines terms for annotating EEG events, including artifacts. This study developed a HED library schema for SCORE, making the terms machine-readable. We demonstrate that the HED-SCORE library schema can be used to annotate events in EEG data stored in the Brain Imaging Data Structure (BIDS). Clinicians and researchers worldwide can now use the HED-SCORE library schema to annotate and compute on electrophysiological data obtained from the human brain.

## *Introduction*

Electroencephalography (EEG) recordings capture brain electrical activity in many different environments. In the clinical setting, a clinician reviews the recordings and, typically using free text, annotates events of interest to identify normal and abnormal findings. Currently, visual analysis and interpretation of the EEG data are critical for diagnosis, and agreement between experts interpreting the same data is higher when they use standardized terminology [1-3]. To facilitate consistency, the International League Against Epilepsy (ILAE) and the International Federation of Clinical Neurophysiology (IFCN) have provided terms and definitions to describe data features in EEG recordings. Here, we incorporate a large set of terms in a system of structured fields following an open science standard, such that these terms can be used in a machine-readable and actionable format.

Recent advances in the Open Science community have developed systems to share definitions of terms for event annotations in order to make these definitions available to a broad user base. The Hierarchical Event Descriptors (HED) system (https://www.hedtags.org/, https://doi.org/10.5281/zenodo.7930927) enables researchers to annotate their data using a formally specified framework for describing, annotating, and validating events occurring during the recording of time-series data [4]. In the HED framework, events are described using tags from a hierarchy of terms. Each event annotation consists of a comma-separated string of terms drawn from the HED vocabulary. This structure allows detailed annotation of events, machine validation of these annotations, and event description-based search across data collected in any number of studies.

The HED framework, vocabulary, and tools have evolved significantly over the past few years [5-7] in alignment with FAIR (Findable, Accessible, Interoperable, and Reusable) guiding principles [8]. HED enables researchers to share, search, and reuse annotated time series data across studies and projects, thereby promoting collaborations and facilitating meta-analyses. The Brain Imaging Data Structure (BIDS) (https://bids.neuroimaging.io) is a growing set of widely used specifications for magnetic resonance imaging (MRI), EEG, intracranial EEG (iEEG), magnetoencephalography (MEG) and positron emission tomography (PET) datasets [9-12]. Incorporating HED annotation into BIDS data formatting facilitates large-scale data-sharing and analysis.

One of the key features of the HED framework is its extensibility. HED allows researchers to add new event types, attributes, and values as needed through schema libraries [4,13]. Schema libraries extend the standard HED schema with structured vocabularies, including terms unique to specific research fields.

To characterize normal as well as pathological (ictal) EEG data using standardized terms, Beniczky et al. [14] presented the Standardized Computer-based Organized Reporting of EEG (SCORE). SCORE aims to improve EEG reporting quality and facilitate data sharing and collaborative research by providing standardized terms and definitions that researchers and clinicians use. SCORE is based on broad international consensus in defining graphoelements, distinct waveforms identified in EEG recordings having unique characteristics. The first SCORE version was endorsed by ILAE-Europe, and the second SCORE version was endorsed as a reporting guideline by the International Federation of Clinical Neurophysiology (IFCN) [15]. SCORE aims to give experts standard terminology that can be used in clinical practice to annotate EEG data and maximize interobserver agreement. SCORE was built on multiple previously proposed guidelines and vocabularies [16-24] and on influential EEG textbooks [25,26]. The second SCORE version added terms drawn from numerous classifications, glossaries, and standard terminologies [19,27-32].

Since its publication, SCORE has been adopted and implemented in many clinical sites and EEG research projects. These studies have used SCORE to better document and study normal and abnormal EEG patterns in different neurological conditions [33-43]. Other studies have investigated the user experience and acceptability of SCORE, reporting that using SCORE has improved the consistency and accuracy of EEG reports [44-47].

To allow the broad scientific and clinical community to add EEG event annotations to BIDS-formatted EEG data in a standardized fashion, we developed a HED-SCORE library schema. Figure 1 shows a schematic view of how a few terms from the extensive HED-SCORE library schema can be used to annotate graphoelements, including artifacts and seizure activity occurring in an electrophysiology time series recording. We further show that HED-SCORE annotations can be used and validated in public BIDS examples.

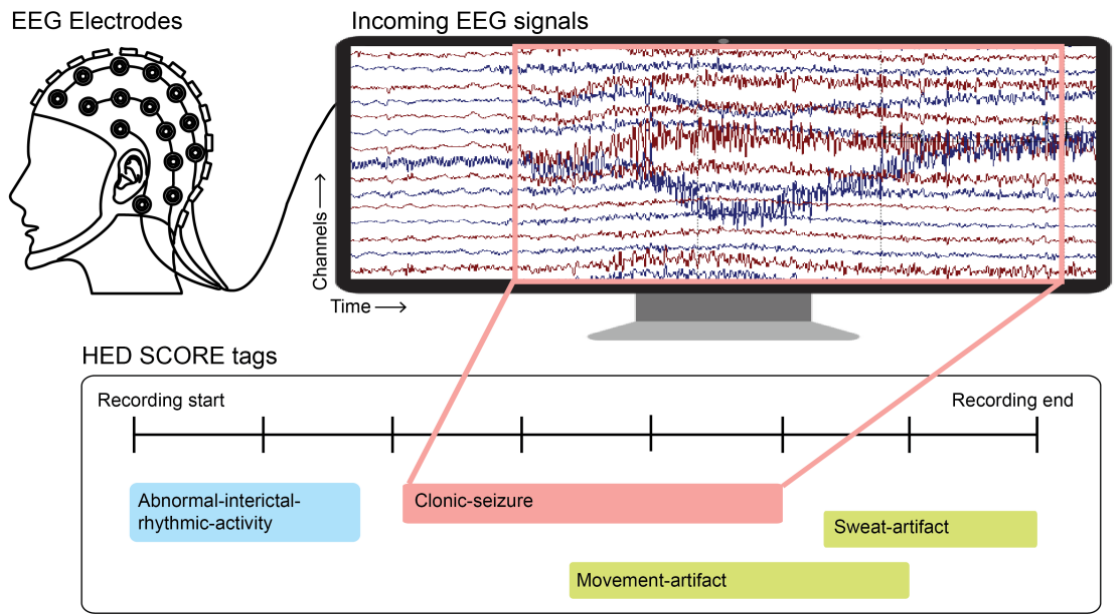

Figure 1 - Schematic of the HED-SCORE library schema implementation. EEG signals are recorded from the scalp and visualized on a computer screen. The signals are visualized and can be annotated using HED-SCORE tags. These tags can be given for, e.g., interictal activity, seizures, and artifacts. Note that this is a schematic, and these features do not correspond to data.

## *Results*

We developed a HED library schema for SCORE. This library schema is an extension of the standard HED schema and allows neurology researchers to annotate electrophysiology recordings per international standards with 396 unique additional terms. An interactive view of the HED-SCORE library schema is available through an expandable HTML viewer (https://www.hedtags.org/display_hed.html). The HTML viewer shows the hierarchical structure of the HED-SCORE library schema (Figure 2). The top level shows the main types of EEG graphoelements (Figure 2A). Each level can be expanded to show the lower-level nodes, such as shown for interictal activity. As the artifact terms are relevant across many neuroimaging fields, these were integrated in the partnered main HED schema, which can be used in conjunction (Figure 2B). Hovering over each tag displays the description of the SCORE term (Figure 2C).

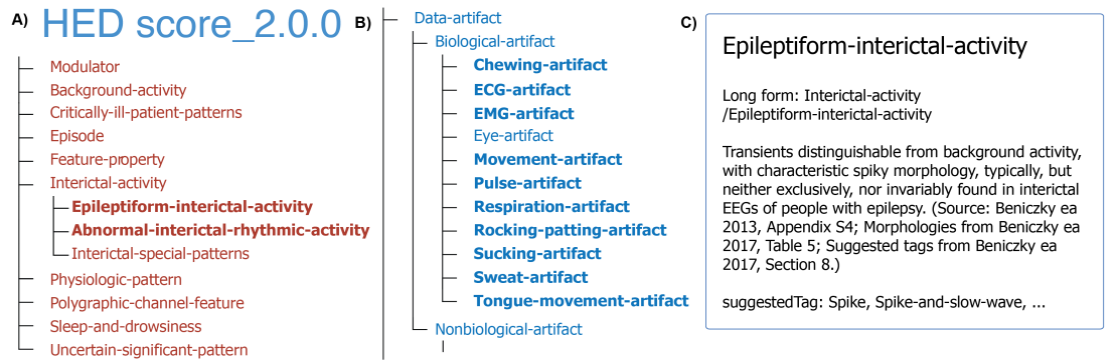

Figure 2 – HED library schema expandable HTML browser allows users to browse the schema terms and descriptors. A) HED-SCORE library schema top-level describing the main types of EEG graphoelements included in SCORE. B) Top-level term '*Artifact*' is expanded to show the lower-level nodes of '*Biological-artifact*'. For data annotations, short forms are typically used (e.g., '*Chewing-artifact*'), and each short form can easily be mapped to its long-form paths ('*Property/Data-property/Data-artifact/Biological-artifact/Chewing-artifact*'). C) The extended description for '*Epileptiform-interictal-activity*' node shown when hovering over the term. The '*suggestedTag*' section shows recommend additional terms that a user might want to include along with this tag.

## HED-SCORE implementation in example BIDS datasets

We provide a set of BIDS examples to test the implementation and validation of HED-SCORE in BIDS (https://github.com/bids-standard/bids-examples/tree/master/xeeg_hed_score ). These examples show the HED-SCORE implementation in three settings: annotating seizures, artifacts, and modulators. Examples are based on data from the Temple University Hospital EEG Corpus (TUEG) [48], an open-source collection of clinical EEG recordings performed at Temple University Hospital (TUH) and Mayo Clinic Rochester, MN. In BIDS, each data file can be accompanied by a metadata file that describes events occurring at certain times during the corresponding recording. When using HED-SCORE annotations in BIDS events files, the first step is to define the HED-SCORE library schema at the project level in the dataset description json file. The BIDS dataset description file must have a field with the HED-SCORE library schema and version (e.g., "HEDVersion": "score_2.0.0"). As the HED-SCORE library schema is partnered with the standard HED schema, any annotations in the data can now include tags from both the HED-SCORE library schema and the standard schema. This avoids duplication of terms and tags from the standard schema, such as '*Left*' and '*Right*', can be used directly to annotate events with HED-SCORE in BIDS.

After the schema is defined in the BIDS dataset description, the second step is to define the HED-SCORE tags that are added to BIDS events files. BIDS events files (…_*events.tsv*) are tab-separated files that can be accompanied by a human and machine-readable field-value JSON sidecar (…_*events.json*). Per BIDS principles, the JSON sidecar describes the columns in the tab separated file and defines the HED-SCORE tags associated with event types in these columns. In this step, the schema is browsed (Figure 2) and relevant terms are defined (examples provided in Figures 3D, 4 and 5). In the third step, data can now be tagged with the terms defined in the JSON sidecar. In the fourth and last step, the HED-SCORE tags are validated using the online HED validator (https://pypi.org/project/hedtools/), which can be used separately or as part the BIDS online validator (https://bids-standard.github.io/bids-validator/). The validators throw errors when a typo is made in the HED-SCORE tags.

### *HED-SCORE can be used to annotate seizure information in BIDS data*

SCORE includes many terms to annotate EEG data in the clinical setting of epilepsy. To test whether HED-SCORE can be used for BIDS data that fit this use-case, the HED-SCORE BIDS example set includes one subject where seizures are annotated (sub-eegSeizureTUH). This example is based on one subject (41, male) from the TUH EEG Seizure Corpus [49]. The TUH Corpus uses 27 labels [50] that were matched to their corresponding HED-SCORE tags (Supplemental Table 1). Figure 3 shows a subset of seizure annotations in the BIDS events file (…_*events.tsv*) where tonic clonic seizure activity spread through different EEG channels listed in the channel column. The accompanying JSON sidecar (…_*events.json*) is used to note how the abbreviation describes the HED-SCORE tag for a tonic-clonic seizure. The sidecar is fully validated in BIDS and if a typing error is made to describe the topic clonic seizure, the BIDS validator throws an error.

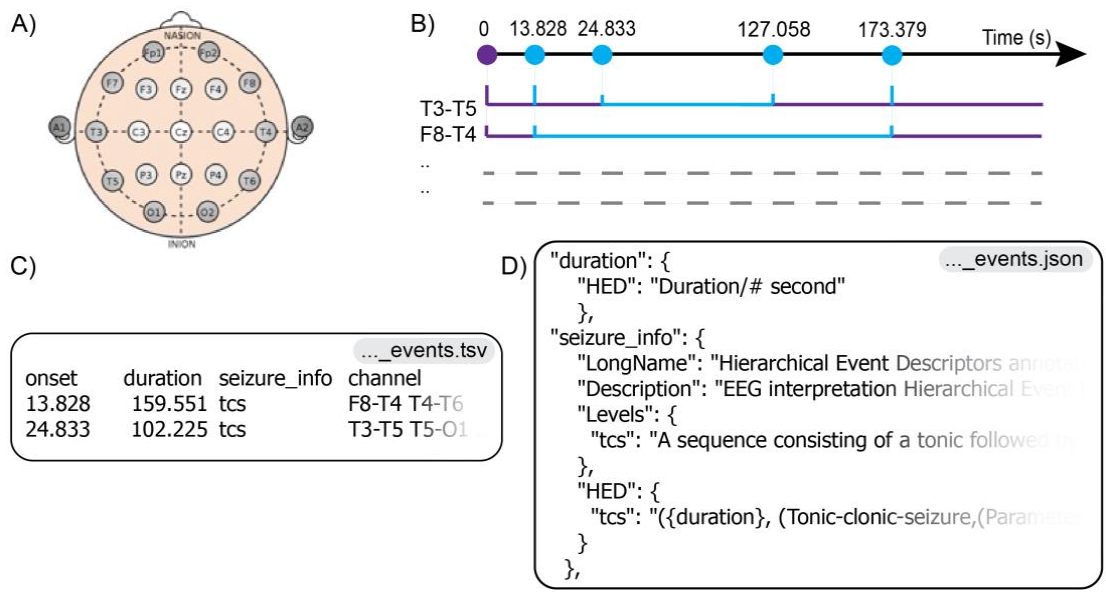

Figure 3 – Schematic example of a section of a BIDS events file (…_events.tsv) with HED-SCORE annotations. A) Electrode location and channel labels of the TUH EEG Seizure Corpus. B) Temporal representation of the example annotations on a subset of selected channels. C) Two example annotations as they appear in the BIDS tab-separated value events file with columns for onset and duration in seconds and a HED column to annotate the seizure information. The column 'channel' is added to indicate which channels in the BIDS _channels.tsv file the annotation corresponds to. The corresponding _channels.tsv file (not visible here, but can be seen in the online example) contains the montage in the dataset. D) An excerpt from the accompanying JSON sidecar describes the HED-SCORE tag used in the column with seizure information (seizure_info).

### *HED can be used to annotate artifacts in BIDS data*

SCORE includes several standardized terms to annotate data artifacts from biological and non-biological sources. The annotation of artifacts is relevant for many neuroimaging data types in BIDS, and artifacts were incorporated in the partnered main HED schema (Figure 2B). The HED-SCORE BIDS example set includes a subject where artifacts are annotated. This example is based on one subject (55, female) from the TUH EEG Artifact Corpus [51], a subset of TUEG that contains different artifacts. The artifacts were matched to HED tags (Supplemental Table 1). The dataset includes annotations of 3 different types of artifacts: eye movements, electrode artifacts, and muscle artifacts. An example of how these artifact annotations are described in the JSON sidecar for the BIDS events file is shown in Figure 4.

HED-SCORE library schema

```
{                                                            ..._events.json
  "onset": {
    "Description": "REQUIRED. Onset (in seconds) of the event, measured from the beginning of the acquisition"
  },
  "duration": {
    "Description": "REQUIRED. Duration of the event (measured from onset) in seconds"
  },
  "annotation_type": {
    "LongName": "Hierarchical Event Descriptors annotations",
    "Description": "EEG interpretation Hierarchical Event Descriptors annotations",
    "Levels": {
      "eyem": "Eye movements that cause artifacts on the recordings.",
      "elec": "Non biological artifacts encompass various electrode related artifacts, such as electrode pop and electrode movement.",
      "musc": "EMG artifact. Artifact from muscle activity and myogenic potentials at the measurements site.",
      "eyem_musc": "Eye and Muscle artifacts."
    },
    "HED": {
      "eyem"      : "Eye-movement-artifact",
      "elec"      : "Nonbiological-artifact",
      "musc"      : "EMG-artifact",
      "eyem_musc" : "(Eye-movement-artifact , EMG-artifact)"
    }
  },
  "channel": {
    "LongName": "Annotated channel",
    "Description": "Space separated list of annotated channels corresponding to *_channels.tsv.",
    "Delimiter": " ",
    "HED": "Parameter-value/#"
  }
}
```

Figure 4 - example of an accompanying events JSON sidecar that lists the HED tags used in annotating the artifacts example (sub-eegArtifactTUH). Required BIDS columns are onset and duration. When annotating HED-SCORE events, the column annotation_type is added to describe with levels and HED tags are used. The column 'channel' is added to describe which channels in the BIDS _channels.tsv file the annotation corresponds to. This _channels.tsv file contains the montage in the dataset.

### *HED-SCORE can be used to annotate modulators in BIDS data*

During EEG recordings for epilepsy, it is common that data can be modulated by effects of medication, eye closure, cognitive tasks or specific manipulations such as hyperventilation or photic stimulation. The HED-SCORE BIDS example set includes a subject where iEEG data were recorded during photic stimulation. This example is based on one subject (46, male) monitored at Mayo Clinic (Rochester, MN) in the Epilepsy Monitoring Unit. The subject provided informed consent, and the research was performed in accordance with the Mayo Clinic Institutional Review Board. It is important to note that the SCORE terminology is intended for scalp EEG, not intracranial EEG (iEEG). Many terms in SCORE don't translate to iEEG due to, e.g., differences in signal properties and standard scalp electrode locations with EEG (i.e., 10-20 system). However, some HED-SCORE tags may be used cautiously in iEEG settings, such as the modulators shown in this example. This example includes photic stimulus procedure annotations where event types and photic stimulation frequencies were annotated in the BIDS events file (…_*events.tsv*) (Figure 5).

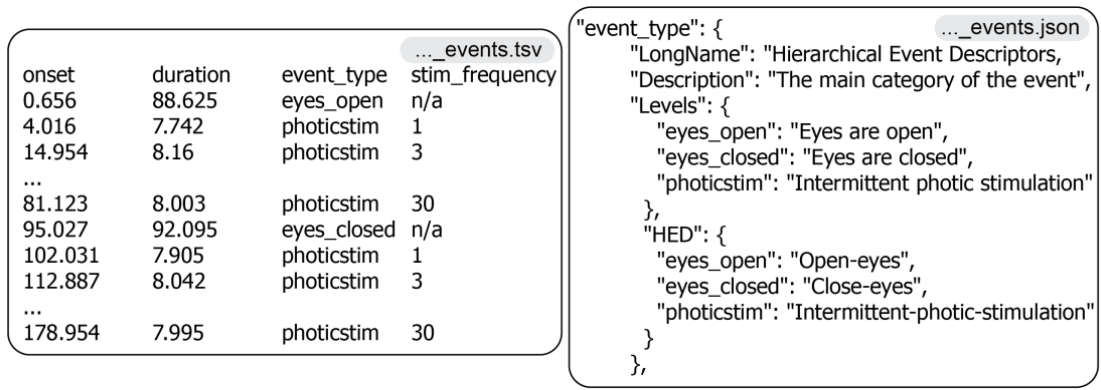

Figure 5 - example of BIDS events file and JSON sidecar excerpts where the HED-SCORE library schema was used to annotate modulators. The events file (left) shows required BIDS onset and duration columns in seconds and an "event_type" column that combines HED-SCORE and HED tags to describe when the subject opened and closed their eyes and when photic stimulation occurred. The JSON sidecar (right) describes the HED-SCORE and HED tags and is necessary for correct validation of the annotations.

## *Discussion*

We implemented SCORE in a HED library schema, such that the SCORE terms are described in structured fields in a human and machine-readable format. This implementation adheres to the HED design principles of uniqueness, clarity, structural sparsity, and orthogonality. The HED-SCORE library schema is compatible with annotating events in BIDS format, and we show several examples of HED-SCORE annotations integrated with the BIDS metadata standards. The HED-SCORE library schema is available in several formats on GitHub and can be viewed in an expandable HTML viewer.

Demonstrating BIDS compatibility is essential for several large efforts where EEG data are shared in BIDS, such as the EEGManyLabs project[52], and open repositories, including OpenNeuro [53], the Cuban Human Brain Mapping Project[54] and data sharing platforms such as LORIS[55]. The HED-SCORE library schema includes not only clinical EEG terms but also terminology that can be used for EEG signal preprocessing, such as artifact annotations. A dataset enriched with the HED-SCORE library schema can be validated using the BIDS validator (https://github.com/bids-standard/bids-

validator) or the HED validator (https://hedtools.org/hed/). This opens an excellent opportunity for researchers to develop BIDS apps[56] and efficiently run automated analysis pipelines with BIDS datasets annotated using the HED-SCORE library schema.

The development of the HED-SCORE library schema is the first step towards developing annotation tools in response to various clinical and scientific needs. For example, CTagger[6] is a user-friendly user interface for easily annotating datasets with HED and the HED-SCORE library schema. EEGLAB[57] tools already incorporate HED and HED-SCORE support, and several initiatives are piloting these integrations, including EEGNet.org (http://eegnet.org/) and the Global Brain Consortium (https://globalbrainconsortium.org/).

Annotating large datasets will be necessary to fully realize the benefits from the open-source implementation of the HED-SCORE library schema. Here, we showed examples of the HED-SCORE library schema applied to existing datasets. Labels used in two datasets from the TUEG archive (https://isip.piconepress.com/projects/tuh_eeg/) were matched to their corresponding HED-SCORE library schema tags. This mapping allows the automatic conversion of the entire TUH data labels to HED-SCORE tags. Moreover, it demonstrates the ease of matching labels in existing annotated databases to their corresponding HED-SCORE library schema tags, and how HED-SCORE can be used to help develop and evaluate automated seizure detection algorithms [58]. The developed HED-SCORE library schema with 396 additional terms provides a base for future tagging efforts by other groups.

SCORE was designed and intended for clinical reports on EEG signals. However, to diagnose and study seizures in individuals with neurological conditions, various methods such as EEG, iEEG, or MEG can be used to capture and record electromagnetic brain activity for both clinical and research purposes. The main distinction between these measurement types is the spatial resolution of the recordings. Where EEG and MEG give a global view, they have a relatively lower spatial resolution, while iEEG has a much higher spatial resolution at the cost of sparse coverage [59,60]. The signals, therefore, have very different features. While some of the scalp EEG graphoelements and findings may apply to iEEG, many do not. Therefore, while some terms from the HED-SCORE library schema may be used for iEEG and MEG data (as in the example with modulators), caution is warranted when describing graphoelements. There are ongoing efforts[60-62] to review and characterize iEEG activity in a standardized manner for clinical purposes, but these have not yet achieved broad consensus.

The HED-SCORE library schema is focused on describing normal and abnormal EEG graphoelements. Therefore, this work does not detail the description of patient information, information related to referral and recording conditions, and administrative data, which are discussed in SCORE papers[14,15]. Moreover, we do not yet include neonatal SCORE templates. The American Clinical Neurophysiology Society standardized EEG terminology and categorization to describe continuous EEG monitoring in neonates[32]. Given the structure of HED, these terms can be included in future extensions of the HED-SCORE library schema or separate library schemas where appropriate.

The HED-SCORE library schema implementation facilitates the sharing of well-annotated data for large-scale analysis. Furthermore, the HED-SCORE library schema can be integrated with existing tools to create new automated analyses, including approaches implemented in future BIDS apps.

## *Methods*

### The HED-SCORE library schema follows HED design principles

The HED-SCORE library schema adheres to HED design principles, basic rules, and form requirements, described in detail in the HED specification (https://hed-specification.readthedocs.io/). Following these principles, the tags are orthogonal, and terms that are used independently are represented in separate hierarchies. The content (library schema) is separate from the presentation and the validation tools. Moreover, every node is unique within the library schema, node names are clear and meaningful on their own. The terms also follow form requirements and individual schema terms begin with a capital letter followed by lowercase letters, and spaces are not allowed. Accordingly, in the case of multiple words, hyphens were used.

## HED-SCORE library schema development

The HED standard schema is organized into subtrees representing aspects of events, such as the type of event, agents involved, actions, items, properties, and relationships between these elements. The HED-SCORE library schema has a top-level organization focused on the identification of EEG features and the top levels of the HED-SCORE library schema correspond to the main types of events described in the SCORE papers [14,15]. These top levels represent background EEG activity, sleep and drowsiness, interictal activity, clinical episodes and electrographic seizures, physiological patterns, patterns of uncertain significance and polygraphic channel features (EMG, EOG, and ECG channels). The top level also includes modulators that pertain to external stimuli and interventions that can change EEG activity. The GitHub commit history reflects the development process of the HED-SCORE library schema, review of internal version HED-SCORE 1.0.0 to current version HED-SCORE 2.0.0 and the design decisions that were made during the development of the HED-SCORE library schema. The pull requests and library schema were reviewed with the HED Working Group, verified against the SCORE papers [14,15] and documented on Zenodo [63].

### *Adhering to orthogonality and avoiding repetition*
In order to adherence to the HED's orthogonality rule and avoid repetition, a *Feature-property* top-level was added to the HED-SCORE library schema. This top-level *Feature-property* is analogous to the HED standard schema *'-Property'*, which includes descriptive elements, such as adjectives and adverbs. These can be used to describe the other tags in more detail. This allows general descriptors to be used alongside several different elements. For example, feature properties include a tag for *Provocative-factor* to annotate one of the modulators that provoked a seizure. In addition, the feature properties include tags for rhythms like '*Gamma-activity*' to describe different types of rhythmic signals like interictal or background activity.

### *Hierarchy follows is-a relationships*
HED requires that child nodes in the schema hierarchy satisfy an 'is-a' relationship with the parent level. This constraint allows users to use 'short-form' annotations and to search general terms. This improves searchability because *Aware-focal-onset-epileptic-seizure* is a type of *Epileptic-seizure*, and a search for *Epileptic-seizure* may return events annotated with *Aware-focal-onset-epileptic-seizure* as well as all other annotated seizure events. HED tools can convert all short-form tags (e.g., *Aware-focal-onset-epileptic-seizure*) to long-form full paths tags (e.g., */Episode/Epileptic-seizure/Focal-onset-epileptic-seizure/Aware-focal-onset-epileptic-seizure*).

### *Partnering with the main HED schema*
Since some of the terms used in SCORE terminology were already present in the standard HED schema (i.e., '*Increasing*, *Decreasing*, '*Symmetrical*', '*Hand*'), these terms were not included in the HED-SCORE library schema. To use these terms, the HED-SCORE library schema was partnered with the standard HED schema. The HED-SCORE library schema is thus designed to not overlap with the standard HED schema and any terms from the main HED schema can be used in conjunction. Moreover, this allowed the tags for artifacts, which apply to many different types of neuroimaging modalities beyond EEG, to be used both using the partnered HED-SCORE library schema and using the main HED schema.

### *Suggested tags to capture interrelated terms*
To maintain the interrelated structure of SCORE while adhering to the HED orthogonality principle (whereby each schema term can be used in only one place) '*suggestedTag*' schema attributes were included for many SCORE terms. Suggested tags indicate other HED tags that should be included when a given term is used to complete the annotation. For example, *Epileptiform-interictal-activity* has suggested morphology tags, such as *Spike* and *Sharp-wave*. Similarly, *Generalized-onset-epileptic-seizure* includes suggested tags for specific seizure types to follow the 2017 seizure classification [28] such as *Tonic-seizure* and *Myoclonic-seizure*. This guides users to consider only those suggested tags that are recommended.

### *Sources*
The SCORE standard was developed over many years by international working groups [14,15]. Every HED-SCORE tag has a permanent GUID and the exact source table or appendix is indicated in the library schema for proper reference and future compatibility.

## HED-SCORE library schema validation

HED tools can be applied to any library schema. According to the standard HED development process for (library) schema and using online HED tools, the HED-SCORE library schema was first developed in the MediaWiki format, a line-oriented markdown language that is easy to read and edit. It was validated to ensure its compliance with current HED (8.0.0+) requirements. As further HED tools for event validation, search and analysis use XML format, HED tools were then used to convert the MediaWiki schema library to XML format.

## HED-SCORE example dataset validation

Several HED resources are available to validate BIDS datasets with HED tags (https://www.hed-resources.org). The example HED-SCORE event and json files were validated using the HED tools python package (https://pypi.org/project/hedtools/). The BIDS HED-SCORE-example dataset was validated using the online BIDS validator (https://bids-standard.github.io/bids-validator/).

**Data Availability**

The BIDS examples (with empty data) are shared on the BIDS-examples page on GitHub (https://github.com/bids-standard/bids-examples/tree/master/xeeg_hed_score).

**Code Availability**

The HED-SCORE library schema is available on GitHub (https://github.com/hed-standard/hed-schemas/tree/main/library_schemas/score); documentation is available on the HED webpage https://hed-schemas.readthedocs.io/en/latest/hed_score_schema.html.

**Acknowledgments**

Research reported in this publication was supported by the National Institute of Mental Health of the National Institutes of Health under Award Numbers R01-MH122258 (DH), and R01-MH126700 (KR, SM). The content is solely the responsibility of the authors and does not necessarily represent the official views of the National Institutes of Health. This was supported in part by the Chengdu MOST grant of 2022 under funding No. GH02-00042-HZ (PVS) and the CNS program of the University of Electronic Sciences and Technology of China (UESTC) under funding No. Y0301902610100201 (PVS).

**Competing interests**

BNL declares intellectual property licensed to Cadence Neuroscience Inc (contractual rights waived; all funds to Mayo Clinic) and Seer Medical Inc (contractual rights waived; all funds to Mayo Clinic), site investigator (Medtronic EPAS, Neuroelectrics tDCS for Epilepsy), and industry consultant (Epiminder, Medtronic, Neuropace, Philips Neuro; all funds to Mayo Clinic).

**Abbreviations**

| | |
|---|---|
| HED | Hierarchical Event Descriptors |
| SCORE | Standardized Computer-based Organized Reporting of EEG |
| BIDS | Brain Imaging Data Structure |
| EEG | Electroencephalography |
| ILAE | International League Against Epilepsy |
| IFCN | International Federation of Clinical Neurophysiology |
| FAIR | Findable, Accessible, Interoperable, and Reusable |
| MRI | Magnetic Resonance Imaging |
| MEG | Magnetoencephalography |
| iEEG | Intracranial Electroencephalography |
| HTML | HyperText Markup Language |
| JSON | JavaScript Object Notation |
| TSV | Tab-separated values |
| XML | Extensible Markup Language |

HED-SCORE library schema

| | |
|---|---|
| TUEG | Temple University Hospital EEG Corpus |
| EMG | Electromyography |
| EOG | Electrooculogram |
| ECG | Electrocardiogram |

HED-SCORE library schema

# Supplementary materials

| Index | TEUG Label | Description (adapted from TUEG [50,51]) | SCORE HED library schema (short-form) |
|---|---|---|---|
| 1 | spsw | Spike and/or slow wave. Patterns of EEGs observed during epileptic seizures. A short duration epileptiform even involving an electrographic spike in activity and/or a slow wave (low frequency wave). Usually no more than 1 sec in duration. | Spike-and-slow-wave |
| 2 | gped | Generalized periodic epileptiform discharge. Periodic diffuse spike/sharp wave discharges across multiple regions or hemispheres. | Generalized-periodic-discharges |
| 3 | pled | Periodic lateral epileptiform discharge. A regular, periodically occurring spike/sharp wave seen in a certain locality of the scalp. | Lateralized-periodic-discharges |
| 4 | eybl | Eyeblink. A specific, sharp, high amplitude eye movement artifact corresponding to blinking of the eye. | Eye-blink-artifact |
| 5 | artf | Artifact. Any non-brain activity electrical signal, such as those due to equipment or environmental factors. | Non-biological-artifact |
| 6 | bckg | Baseline/non-interesting events, all other non-seizure cerebral signals | Not applicable, baseline is the default |
| 7 | seiz | Seizure. A basic annotation for seizures. | Epileptic-seizure |
| 8 | fnsz | Focal nonspecific seizure. A large category of seizures occurring in a specific focality. | Focal-onset-epileptic-seizure |
| 9 | gnsz | Generalized seizure. A large category of seizures occurring in most if not all of the brain. | Generalized-onset-epileptic-seizure |
| 10 | spsz | Simple partial seizure. Brief seizures that start in one location of the brain (and may spread) where the patient is fully aware and able to interact. | Aware-focal-onset-epileptic-seizure |
| 11 | cpsz | Complex partial seizure. Same as simple partial seizure (spsz) but with impaired awareness. | Impaired-awareness-focal-onset-epileptic-seizure |
| 12 | absz | Absence seizure. Brief, sudden seizure involving lapse in attention. | Absence-seizure |
| | | Usually lasts no more than 5 seconds and commonly seen in children. | |
| 13 | tnsz | Tonic seizure. A seizure involving the stiffening of the muscles. Usually associated with and annotated as tonic-clonic seizures, but not always (rarely there is no clonic phase). | Tonic-seizure |
| 14 | cnsz | Clonic seizure. A seizure involving sustained, rhythmic jerking. | Clonic-seizure |
| 15 | tcsz | Tonic-clonic seizure. A seizure involving loss of consciousness and violent muscle contractions. | Tonic-clonic-seizure |
| 16 | atsz | Atonic seizure. A seizure involving the loss of tone of muscles in the body. | Atonic-seizure |
| 17 | mysz | Myoclonic seizure. A seizure associated with brief involuntary twitching or myoclonus. | Myoclonic-seizure |
| 18 | nesz | Non-epileptic seizure. Any non-epileptic seizure observed. Contains no electrographic signs. | Seizure-PNES |
| 19 | intr | Interesting patterns. Any unusual or interesting patterns observed that don't fit into the above classes. | Uncertain-significant-pattern |
| 20 | slow | Slowing. A brief decrease in frequency | Dependent on the case, can use Temporal-slowing-elderly or Feature-frequency/Decreasing |
| 21 | eyem | Eye movement. A very common frontal/prefrontal artifact seen when the eyes move. | Eye-movement-artifact |
| 22 | chew | Chewing. A specific artifact involving multiple channels that corresponds with patient chewing, "bursty". | Chewing-artifact |
| 23 | shiv | Shivers. A specific, sustained sharp artifact that corresponds with patient shivering. | Movement-artifact |
| 24 | musc | Muscle artifact. A very common, high frequency, sharp artifact that corresponds with agitation/nervousness in a patient. | EMG-artifact |
| 25 | elpp | Electrode pop. A short artifact characterized by channels using the same electrode "spiking" with perfect symmetry. | Electrode-pops-artifact |
| 26 | elst | Electrostatic artifact. Artifact caused by movement or interference on the electrodes, variety of morphologies. | Induction-artifact |

HED-SCORE library schema| | | | |
|---|---|---|---|
| **27** | calb | Artifact caused by calibration of the electrodes. Appears as a flattening of the signal in the beginning of files. | Nonbiological-artifact |
| **28** | hphs | Hypnagogic hypersynchrony, a brief period of high amplitude slow waves | Hypnagogic-hypersynchrony |
| **29** | trip | Large, three-phase waves frequently caused by an underlying metabolic condition. | Triphasic-morphology |
| **30** | elec | This tag was developed for the TUAR artifact corpus an indicates electrode artifacts that encompass three electrode related artifacts. 1) Electrode pop is an artifact characterized by channels using the same electrode "spiking" with an electrographic phase reversal. 2) Electrostatic is an artifact caused by movement or interference of electrodes and or the presence of dissimilar metals. 3) A lead artifact is caused by the movement of electrodes from the patient's head and or poor connection of electrodes. This results in disorganized and high amplitude slow waves. | Nonbiological-artifact |

Table 1 – Labels and descriptions used to annotate the Temple University Hospital EEG Corpus (TUEG)[46,47,50,51] and their corresponding SCORE HED library schema annotations. In the TUEG Corpus all channels with baseline activity were tagged as 'bckg'. In the BIDS examples, however, background activity is not tagged, as we assume that the default is that channels show baseline activity and the TUEG 'bckg' label is not translated to an HED-SCORE tag. A direct translation for TUEG label 'slow' was not found in HED-SCORE, but a user can use Temporal-slowing-elderly or Feature-frequency/Decreasing dependent on the situation.

23